\documentclass[twocolumn,pre,floatfix,superscriptaddress, nofootinbib]{revtex4}
\usepackage[T1]{fontenc}
\usepackage[latin9]{inputenc}
\usepackage{mathrsfs}
\usepackage{amsmath}
\usepackage{amssymb}
\usepackage{graphicx}
\usepackage{braket}
\usepackage{color}
\usepackage[svgnames]{xcolor}
\usepackage{comment}
\usepackage{natbib}
\usepackage{pifont} 

\makeatletter

\usepackage{listings}

\usepackage{hyperref}

\hypersetup{
  colorlinks=true,
  linkcolor=blue,
  citecolor=ForestGreen,
  urlcolor=DarkOrchid
}

\newcommand{\TTO}{\mathrm{TTO}}
\newcommand{\TTS}{\mathrm{TTS}}
\newcommand{\HP}{\mathbb{H}}
\newcommand{\MIS}{\mathrm{MIS}}
\newcommand{\codevar}[1]{\texttt{#1}}
\newcommand{\Fib}{\mathrm{Fib}}
\newcommand{\OO}{\mathcal{O}}
\newcommand{\OM}{\OO_{\mathrm{mem}}}
\newcommand{\filling}{\varrho}

\begin{document}

\title{Hardness of the Maximum Independent Set Problem on Unit-Disk Graphs and Prospects for Quantum Speedups}

\author{Ruben S. Andrist}
\thanks{These authors contributed equally.}
\affiliation{Amazon Quantum Solutions Lab, Seattle, Washington 98170, USA}

\author{Martin J. A. Schuetz}
\thanks{These authors contributed equally.}
\affiliation{Amazon Quantum Solutions Lab, Seattle, Washington 98170, USA}
\affiliation{AWS Center for Quantum Computing, Pasadena, CA 91125, USA}

\author{Pierre Minssen}
\thanks{These authors contributed equally.}
\affiliation{Global Technology Applied Research, JPMorgan Chase, New York, NY 10017 USA}

\author{Romina Yalovetzky}
\thanks{These authors contributed equally.}
\affiliation{Global Technology Applied Research, JPMorgan Chase, New York, NY 10017 USA}

\author{Shouvanik Chakrabarti}
\affiliation{Global Technology Applied Research, JPMorgan Chase, New York, NY 10017 USA}

\author{Dylan Herman}
\affiliation{Global Technology Applied Research, JPMorgan Chase, New York, NY 10017 USA}

\author{Niraj Kumar}
\affiliation{Global Technology Applied Research, JPMorgan Chase, New York, NY 10017 USA}

\author{Grant Salton}
\affiliation{Amazon Quantum Solutions Lab, Seattle, Washington 98170, USA}
\affiliation{AWS Center for Quantum Computing, Pasadena, CA 91125, USA}
\affiliation{California Institute of Technology, Pasadena, CA, USA}

\author{Ruslan Shaydulin}
\affiliation{Global Technology Applied Research, JPMorgan Chase, New York, NY 10017 USA}

\author{Yue Sun}
\affiliation{Global Technology Applied Research, JPMorgan Chase, New York, NY 10017 USA}

\author{Marco Pistoia}
\altaffiliation[]{These authors acted as Co-PIs.}
\affiliation{Global Technology Applied Research, JPMorgan Chase, New York, NY 10017 USA}

\author{Helmut G. Katzgraber}
\altaffiliation[]{These authors acted as Co-PIs.}
\affiliation{Amazon Quantum Solutions Lab, Seattle, Washington 98170, USA}

\date{\today}

\begin{abstract}

Rydberg atom arrays are among the leading contenders for the demonstration of
quantum speedups. Motivated by recent experiments with up to 289 qubits
[Ebadi \textit{et al.}, Science \textbf{376}, 1209 (2022)] we study the
maximum independent set problem on unit-disk graphs with a broader range of classical solvers beyond the scope of the original paper.
We carry out extensive numerical studies and assess problem hardness, using
both exact and heuristic algorithms. 
We find that quasi-planar instances with Union-Jack-like connectivity can be
solved to optimality for up to thousands of nodes within minutes, with both
custom and generic commercial solvers on commodity hardware, without any
instance-specific fine-tuning.
We also perform a scaling analysis, showing that by relaxing the constraints on the classical simulated annealing algorithms considered in Ebadi \textit{et al.}, our implementation is competitive with the quantum algorithms. 
Conversely, 
instances with larger connectivity or less structure are shown to display
a time-to-solution potentially orders of magnitudes larger. 
Based on these results we propose protocols to systematically tune problem
hardness, motivating experiments with Rydberg atom arrays on instances orders
of magnitude harder (for established classical solvers) than previously
studied.
\end{abstract}

\maketitle

\section{Introduction}
\label{introduction}

Combinatorial optimization problems are pervasive across
science and industry, with prominent applications in areas such as
transportation and logistics, telecommunications, manufacturing, and finance.
Given its potentially far-reaching impact, the demonstration of quantum speedups
for practically relevant, computationally hard problems (such as combinatorial
optimization problems) has emerged as one of the greatest milestones in quantum
information science.

Over the last few years, programmable Rydberg
atom arrays have emerged as a promising platform for the implementation of
quantum information protocols \citep{wilk:10, saffman:10, saffman:16,
bernien:17, henriet:20, adams:20, ebadi:21}, and (in particular) quantum
optimization algorithms \citep{pichler:18, pichler:18computational, zhou:20,
serret:20, ebadi:22, cain:23, schiffer:23}. 
Some of the exquisite, and experimentally demonstrated capabilities of these
devices include the deterministic positioning of individual neutral atoms in
highly scalable arrays with arbitrary arrangements \citep{endres:16,
barredo:18}, the coherent manipulation of the internal states of these atoms
(including excitation into strongly excited Rydberg states) \citep{labuhn:16,
lienhard:18, guardado:18}, the ability to coherently shuttle around
individual atoms \citep{bluvstein:21}, and strong interactions mediated by
the Rydberg blockade mechanism \citep{lukin:01, ebadi:22, levine:19,
evered:23}.

\begin{figure}
  \includegraphics[width=1.0 \columnwidth]{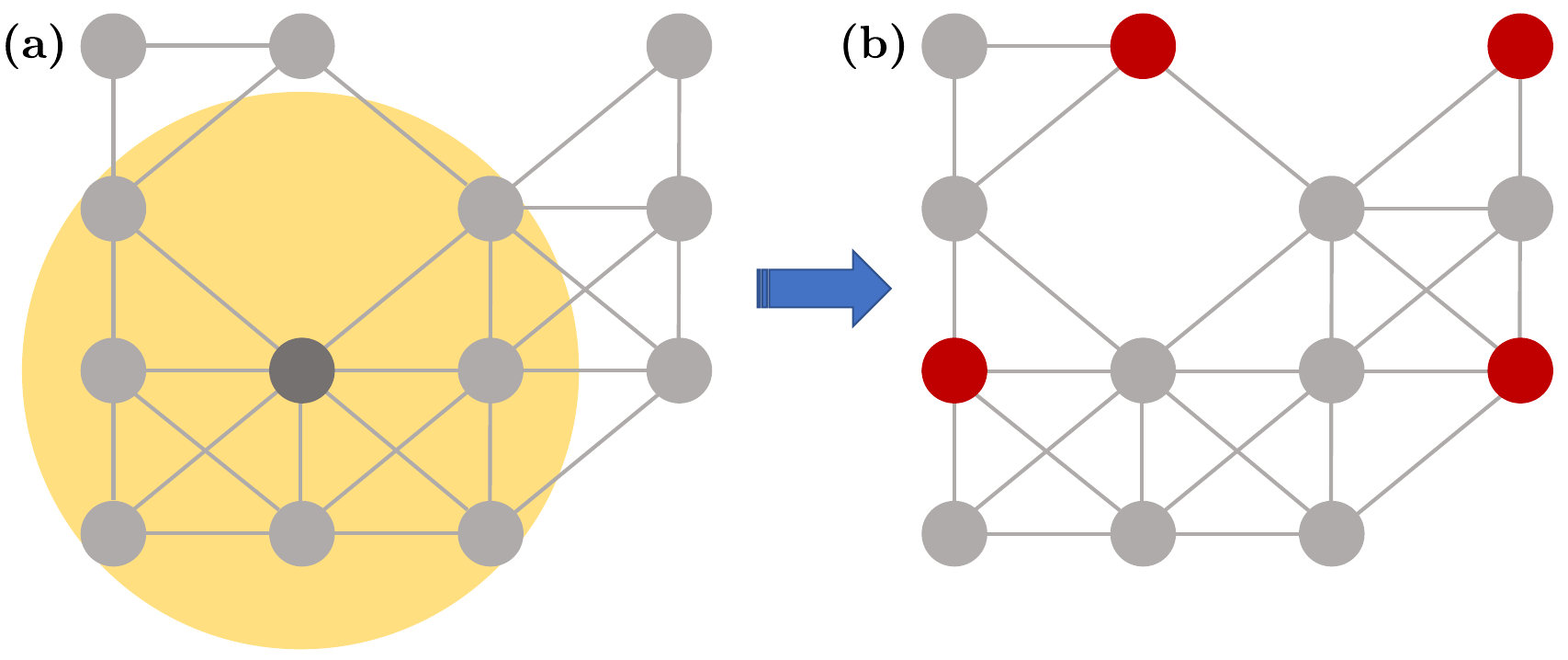}
  \caption{
    Schematic illustration of the problem. 
      \textbf{(a)} We consider unit-disk graphs with nodes arranged on a
      two-dimensional square lattice with lattice spacing $a$ and filling
      fraction $\filling \sim 80\%$, and edges connecting all pairs of nodes
      within a unit distance (illustrated by the circle). For $\sqrt{2}a \leq
      R_{b} < 2a$ (as considered here), nodes are connected to nearest and
      next-nearest neighbors resulting in a (quasi-planar) Union-Jack pattern
      with maximum degree $d_{\mathrm{max}}=8$. 
      \textbf{(b)} Our goal is to solve the MIS problem on this family of
      instances (as depicted here with nodes colored in red in the right panel) and assess the
      hardness thereof using both exact and heuristic algorithms.  
    \label{fig:scheme}
  }
\end{figure}

The physics of the Rydberg blockade
mechanism has been shown to be intimately related to the canonical (NP-hard)
maximum independent set (MIS) problem \citep{pichler:18}, in particular for
unit-disk graphs.
The MIS problem involves finding the largest independent set of vertices in a
graph, i.e., the largest subset of vertices such that no edges connect any pair
in the set; compare Fig.~\ref{fig:scheme} for a schematic illustration. 
As shown in Ref.~\citep{pichler:18}, MIS problems can be encoded with
(effectively two-level) Rydberg atoms placed at the vertices of the target
(problem) graph. Strong Rydberg interactions between atoms then prevent two neighboring atoms from being simultaneously in the excited Rydberg
state, provided they are within the Rydberg blockade radius, thereby
effectively implementing the independence constraint underlying the MIS
problem. By virtue of this Rydberg blockade mechanism, Rydberg
atom arrays allow for a hardware-efficient encoding of the MIS problem on
unit-disk graphs, with the (tunable) disk radius $R_{b} \sim
1$ -- $10\mu{\mathrm{m}}$ setting the relevant length-scale \citep{adams:20}.

\textbf{Overview of main results.} Recently, a potential (superlinear) quantum
speedup over classical simulated annealing has been reported for the MIS
problem \citep{ebadi:22}, based on variational quantum algorithms run on
Rydberg atom arrays with up to 289 qubits arranged in two spatial dimensions.
This work focused on benchmarking quantum variational algorithms against simulated annealing by viewing it as a classical analogue of the adiabatic algorithm, yet left open the question of benchmarking against other state-of-the-art classical solvers.
Motivated by this experiment, we perform a detailed analysis of the MIS problem on
unit-disk graphs and assess problem hardness using both exact and heuristic
methods. We provide a comprehensive algorithmic and numerical analysis, and we
demonstrate the following: 
(i) Typical quasi-planar instances with Union-Jack-like connectivity (as
  studied in Ref.~\citep{ebadi:22}) can be solved to optimality for up to
  thousands of nodes within minutes, with both custom and generic commercial
  solvers on commodity hardware, without any instance-specific fine-tuning.
(ii) Systematic scaling results are provided for all solvers, displaying
  qualitatively better runtime scaling for solvers exploiting the quasi-planar
  problem structure than generic ones. In particular, we find that by relaxing the detailed balance constraint, and considering the low depth regime (both of which are required for analytic runtime lower bounds on SA described in Ref.~\citep{cain:23}) our implementation of classical simulated annealing is competitive with the quantum algorithm's performance in Ref.~\citep{ebadi:22}.
(iii) Conversely, while the definition of problem hardness may be specific to
  the method used, instances with larger connectivity or less structure
  display a time-to-solution typically orders of magnitudes larger. 
(iv) Based on these results, we propose protocols to systematically tune
  problem hardness (as measured by classical time-to-solution), motivating
  experiments with Rydberg atom arrays on instances orders of magnitude harder
  (for established classical solvers) than previously studied. 

This paper is organized as follows. 
In Sec.~\ref{problem} we first formalize the problem we consider. 
Next, in Sec.~\ref{methods} we describe the algorithmic tool suite with which
we address this problem. 
In Sec.~\ref{numerics} we then describe our numerical experiments in detail. 
Finally, in Sec.~\ref{conclusion}, we draw conclusions and give an outlook on
future directions of research.

\section{Problem Specification}
\label{problem}
 
The MIS problem is a prominent combinatorial
optimization problem with practical applications in network design
\citep{hale:80}, vehicle routing \citep{dong:22}, and finance
\citep{boginski:05, kalra:08}, among others, and is closely related to the
maximum clique, minimum vertex cover, and set packing problems
\citep{wurtz:22}. 

\subsection{Definition}
Formally, the MIS problem reads as follows. Given an undirected graph $\mathcal{G}=(\mathcal{V}, \mathcal{E})$, an independent set $S\subseteq\mathcal{V}$ is a subset of vertices of $\mathcal{G}$ such that no two vertices in $S$ share an edge in $\mathcal{E}$. The maximum independent set problem is then the task of finding the \textit{largest} independent set in $\mathcal{V}$. The cardinality of this largest independent set is referred to as the \textit{independence number} $|\mathrm{MIS}|$.
One way to formulate the MIS problem mathematically is to first associate
a binary variable $x_{i} \in \{0,1\}$
to every vertex $i \in \mathcal{V}$, such that $x_{i}=1$ if vertex $i=1, \dots, N$
belongs to the independent set, and $x_{i}=0$ otherwise. The MIS problem can
then be expressed as a compact integer linear program of the form 
\begin{equation}
  \begin{aligned}
    \max \quad & \sum_{i} x_{i} \\
    \textrm{s.t.} \quad & x_{i} + x_{j} \leq 1 \quad \forall (i,j) \in \mathcal{E}\\
    & x_{i} \in \{0,1\} \quad i=1, \dots, N,   \\
  \end{aligned}
  \label{eq:ip}
\end{equation}
with the objective to maximize the marked vertices while adhering to the
independence constraint. Generalizations to the maximum-weight independent set
problem are straightforward \citep{dong:22}. 

A formulation of the MIS problem that is commonly used in the physics literature expresses the integer linear program in Eq.~\eqref{eq:ip} in terms of a Hamiltonian that includes a (soft) penalty to
non-independent configurations (i.e., when two vertices in the set are connected by
an edge) \citep{pichler:18}. This Hamiltonian is given by  
\begin{equation}
H = -\sum_{i} x_{i} + V \sum_{(i,j) \in \mathcal{E}} x_{i}x_{j},
\label{eq:hamiltonian}
\end{equation}
with a negative sign in front of the first term because the largest independent
set is searched for within a minimization problem, and where the penalty parameter
$V$ enforces the constraints. Energetically, this Hamiltonian favors having each
variable in the state $x_{i}=1$ unless a pair of vertices are connected by
an edge. For $V>1$, the ground state is guaranteed to be a MIS, because it is strictly more favorable to have at most one vertex per edge in the set as opposed to both vertices being marked \citep{ebadi:22}. 
Still, within this framework, the independence constraint typically
needs to be enforced via post-processing routines (as is done, for example,
in Ref.~\citep{ebadi:22}).
Mapping the binary variables $x_{i}$ to two-level Rydberg atoms subject to a
coherent drive with Rabi frequency $\Omega$ and detuning $\Delta$, one can then
search for the ground state of the Hamiltonian $H$ (encoding the MIS) via, for
example, quantum-annealing-type approaches using quantum tunneling between
different spin configurations \citep{pichler:18, ebadi:22}. 

\subsection{Problem hardness} 

The MIS problem is known to be strongly NP-hard,
making the existence of an efficient algorithm for finding the maximum
independent set on generic graphs unlikely.  As such, the MIS problem is
even hard to approximate \citep{garey:90}, and in general cannot be approximated to a constant factor in polynomial time (unless P = NP). 

Here, however, we primarily
focus on the MIS problem on unit-disk graphs (dubbed MIS-UD hereafter), given
their intimate relation to Rydberg physics \citep{pichler:18, ebadi:22}. 
Our main goal is to empirically assess the hardness of MIS-UD.
As schematically depicted in Fig.~\ref{fig:scheme}, unit-disk graphs are
defined by vertices on a two-dimensional plane with edges connecting all pairs
of vertices within a unit distance. The MIS-UD problem appears in practical
situations with geometric constraints such as map labeling \citep{agarwal:98}
and wireless network design \citep{hale:80}. While approximate solutions to
MIS-UD can be found in polynomial time \citep{leeuwen:05}, solving the problem
\textit{exactly} is still known to be NP-hard for worst-case instances
\citep{ebadi:22, clark:90}. 

\subsection{Problem instances and figures of merit} 

The unit disk (UD) 
problem instances of interest can be
characterized by the number of nodes $N$, the side length of the underlying
square lattice $L$, and the filling fraction $\filling$, with $N \approx \filling
L^{2}$, as schematically depicted in Fig.~\ref{fig:scheme}. 
Following Ref.~\citep{ebadi:22}, we focus on single-component (non-planar,
but quasi-planar) UD instances with nearest and next-nearest (diagonal)
couplings only, resulting in Union-Jack-type graphs with maximum degree
$d_{\mathrm{max}}=8$. 
Accordingly, these instances consist of (at maximum) $n_{\mathrm{corner}} \leq
4$ \textit{corner}, $n_{\mathrm{boundary}} \leq 4(L-2)$ \textit{boundary}, and
$n_{\mathrm{bulk}} \leq (L-2)^{2}$ \textit{bulk} nodes, with (at maximum)
three, five, and eight neighbors, respectively, and a total of at most
$|\mathcal{E}|_{\mathrm{max}}$ edges, with
$|\mathcal{E}|_{\mathrm{max}}=4L^2-6L+2$. 
Because $|\mathcal{E}|_{\mathrm{max}} \sim N$, the graph density $D_{\mathrm{graph}}$ scales as $D_{\mathrm{graph}} = 2|\mathcal{E}|_{\mathrm{max}}/N(N-1) \sim 1/N$, showing that
these instances become sparser as the system size $N$ grows. 
If not otherwise specified, we take $\filling=80\%$, as was done in Ref.~\citep{ebadi:22}. 
For comparison, we also run experiments on (unstructured) random
Erd\H{o}s-R\'enyi (ER) graphs denoted as $\mathcal{G}(n,m)$, chosen uniformly at
random from the collection of all graphs with $n$ nodes and $m$ edges, or
similarly $\mathcal{G}(n,p)$ for graphs constructed by connecting nodes
randomly with probability $p$.  

To assess and compare the performance of various
algorithms (as specified below) we consider the following figures of merit. 
We use $|\mathrm{MIS}|$ to denote the independence number, while $P_{\mathrm{MIS}}$ refers to the probability of observing an (exact) MIS within a fixed number of steps \citep{ebadi:22}. 
For a given instance, many MIS solutions may be available, with the
corresponding number of MIS solutions (i.e., the MIS degeneracy) denoted as
$D_{\mathrm{MIS}}$. Similarly, the quantity $D_{\mathrm{MIS-1}}$ refers to the
number of first excited states (i.e., independent sets of size
$|\mathrm{MIS}|-1$). 
As shown in Ref.~\citep{ebadi:22}, in the context of simulated annealing, problem
hardness may further be specified in terms of the conductance-like hardness
parameter
\begin{equation}
\HP =\frac{D_{\mathrm{MIS-1}}}{|\mathrm{MIS}|\cdot D_{\mathrm{MIS}}},
\label{eq:hardness}
\end{equation}
with the factor $|\mathrm{MIS}|D_{\mathrm{MIS}}$ denoting the number of
possible transitions from a first excited state into a MIS ground state. 
Finally, we are interested in the \textit{time-to-solution} (TTS). 
For exact methods, TTS refers to the time needed to find the optimal solution
(i.e., ground state). While the optimum may be found after time TTS, additional
time may be required to provide an optimality certificate, resulting in the
\textit{time-to-optimality} (TTO) time scale, with $\mathrm{TTO} \geq \mathrm{TTS}$. 
By definition, provable optimality is not available with heuristic methods.
Here, we define $\mathrm{TTS}_{99}$ as the time required to find the exact solution (ground
state) with 99\% success probability. We can then write $\mathrm{TTS}_{99}$ as  
\begin{equation}
\mathrm{TTS}_{99} = \tau R_{99},
\end{equation}
where $\tau$ refers to the time of a single run (shot) and 
\begin{equation}
R_{99}=\frac{\log(1-0.99)}{\log(1-P_{\mathrm{MIS}})}
\end{equation}
is the number of shots (repetitions) needed to reach the desired success
probability \citep{aramon:19}. For small values of $P_{\mathrm{MIS}}$, we have
$R_{99} \approx 4.6/P_{\mathrm{MIS}}$, showing that the success probability of
a \textit{single} run $P_{\mathrm{MIS}}$ determines the inverse of the time-to-solution for
heuristic algorithms.

\section{Algorithmic Tool Suite}
\label{methods}

In this section we detail the algorithms used to solve the MIS problem on
UD graphs. We distinguish between \textit{exact} methods (which by design can
deterministically find the ground state, typically at the expense of an
exponential runtime) and \textit{heuristics} (which cannot provide an
optimality certificate, but may require shorter runtimes).

\subsection{Exact Methods}

\begin{figure}
  \includegraphics[width=1.0 \columnwidth]{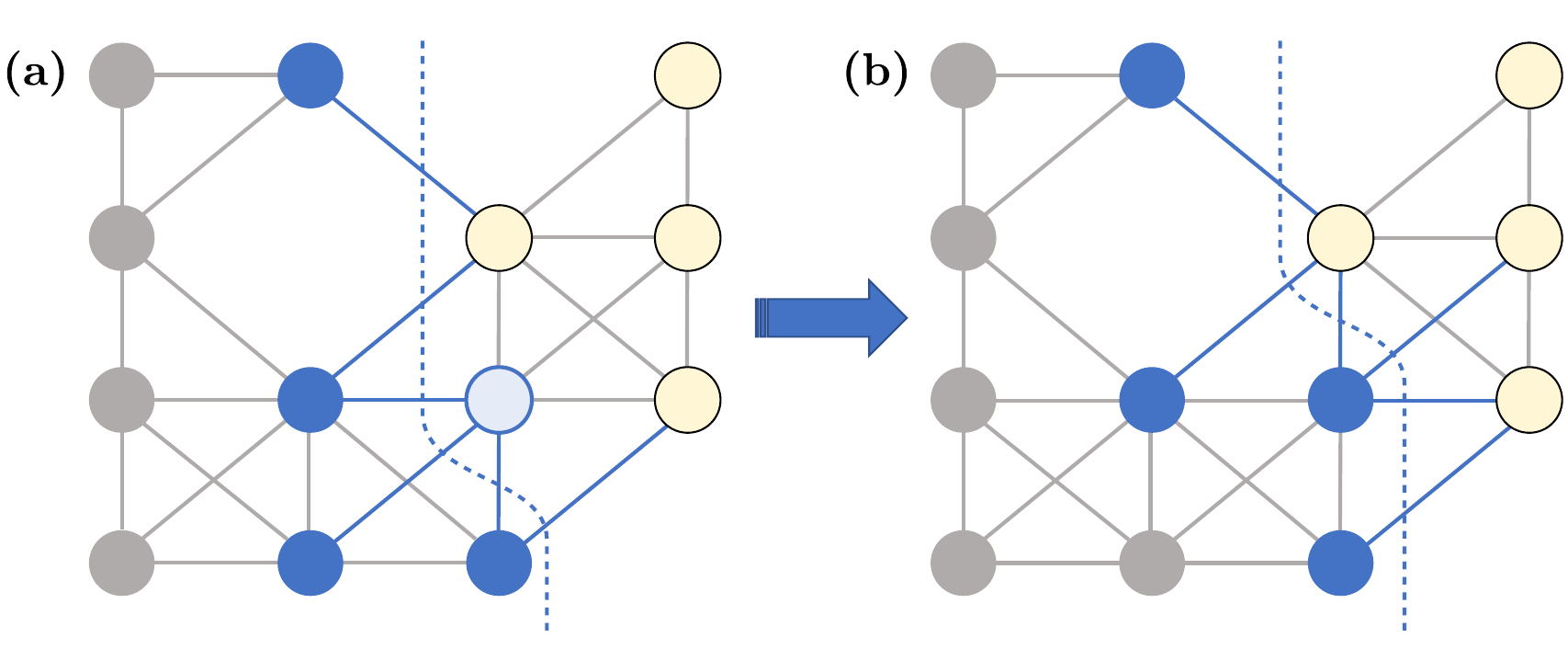}
  \caption{
    Schematic illustration of the (exact) sweeping line algorithm (SLA) as
    applied to the MIS-UD problem. \textbf{(a)} SLA proceeds by sweeping a fictitious
    line across the graph and tracking all potentially optimal MIS
    configurations on this boundary, efficiently exploiting the quasi-planar
    structure of the UD graph. Processed nodes are shown in gray, boundary
    nodes in blue and unprocessed nodes in yellow. \textbf{(b)} The light
    blue node from (a) is added to the boundary, while the bottom left blue
    node is dropped (as a result of not having any more connections to
    unprocessed nodes).  
  \label{fig:sla}}
\end{figure}

\textbf{Sweeping line algorithm.} We first consider an exact sweeping line
algorithm (SLA) \citep{shamos:76} that efficiently exploits the quasi-planar structure of the UD
instances considered here. The anatomy of the SLA is schematically illustrated
in Fig.~\ref{fig:sla}. The SLA is based on full enumeration and
works by sweeping a fictitious line across the two-dimensional plane (e.g.,
from left to right). Specifically, the algorithm proceeds as follows. 
We define the \textit{boundary} as the set of all processed nodes which still
share an edge with an unprocessed node. At each step, we track the
size of the largest independent set (so far) for each valid boundary variant
(set of assigned nodes on the boundary).
As the line is swept across the graph, it stops at every node $i=0, \dots, N-1$.
We then generate the new variants at step $i$ from those at step $i-1$ as
follows:
\begin{itemize}
  \item If the variant has a neighboring node of $i$ assigned, we generate
    only a new variant with $i$ not selected (that is, $x_{i}=0$).
  \item Otherwise we also create a new variant with $i$ assigned as $x_{i}=1$
    (which increases its independent set).
\end{itemize}
Note that forward-looking information is not required, and only boundary nodes
are relevant for this decision. Once the new variants have been generated, node
$i$ becomes part of the boundary, and we proceed with the next step. When moved
across the graph, this recipe generates all valid sets with runtime
$\mathcal{O}(2^N)$. However, we can efficiently summarize information that is
not relevant to finding the MIS:
\begin{itemize}
  \item Adding new nodes typically removes older nodes from the boundary (because
    they no longer have a connection to an unprocessed node); c.f.
    Fig.~\ref{fig:sla}.
  \item We only need to track the size of the largest independent set for each
    boundary variant (i.e., for any boundary configuration we can discard any
    option with equal or smaller number of assignments).
\end{itemize}
For the UD graphs considered here, the number of variants tracked on any given
boundary is limited by the number of valid assignments on the
boundary, $\#\MIS$. When processing the nodes in order, the boundaries form continuous
(mostly) one-dimensional strips and $\#\MIS(L) \leq \Fib(L+1)$ (which can be
shown by induction). As a result, the memory requirement for SLA scales as
$\OM(\Fib(L)) = \OM(\Fib(\sqrt{N}))$ (to hold the variants at each step).
SLA finds the optimal solution after $N$ steps ($\TTS=\TTO$), processing all
variants at each step with a runtime of $\OO(N\Fib(L)) = \OO(N\Fib(\sqrt{N}))
\approx \OO(N \varphi^{\sqrt{N}})$, where $\varphi \approx 1.62$ is the golden ratio.
This procedure can also be modified to count the degeneracy of the ground and
first excited state (i.e., $D_{\mathrm{MIS}}$ and $D_{\mathrm{MIS}-1}$,
respectively) by adjusting summarization accordingly.
  
\textbf{Branch \& bound solvers.} We complement our custom SLA solver with commercial solvers based on the branch and bound (B\&B) search method. In particular, we use the solver offered by CPLEX \cite{cplex2021v20.1}; similar results were observed with Gurobi \cite{gurobi}.
In practice, these solvers are among the de facto go-to tools for many hard,
mixed-integer optimization problems. 
By design, B\&B solvers provide upper and lower bounds on the solution, with
the difference between these yielding an \textit{optimality gap}, thereby
giving information about the quality of the solution (at any step throughout
the algorithmic evolution). Assuming a maximization problem, the lower bound
corresponds to the best known feasible solution whereas the upper bound refers
to the optimal value for the corresponding relaxed problem in the B\&B
procedure.
In this work we focus on the TTS and TTO figures of merit, which are readily
provided by our chosen solvers.
We evaluate TTO by enforcing a zero gap between the upper and the lower bound,
and TTS by setting the upper bound to be the optimal solution found
previously. Thus, the solver terminates successfully as soon as it reaches
the optimal solution. Typically, we find $\mathrm{TTO}$ is strictly greater than $\mathrm{TTS}$ because of additional time required to prove optimality. 
In order to draw a clear line between the B\&B solvers and the heuristic solvers described 
below, we deactivate the B\&B solvers' capability to find feasible
solutions heuristically. 
As such, in practice we expect smaller values for TTS when utilizing these
additional features of modern B\&B solvers, effectively making the TTS values we
report here upper \textit{bounds} for B\&B based performance. 
Finally, to account for
the multi-threading capabilities of B\&B solvers, we report the \textit{process time}, i.e., the sum of system and user CPU seconds of each core used during the calculation, rather than the \textit{wall-clock time}. 
We checked that the multi-threading overhead does not impact the $\mathcal{O}$ complexities inferred from the numerical experiments; see Appendix \ref{classical-hardware} for more details.

\subsection{Heuristics}

\begin{figure}
  \includegraphics[width=1.0 \columnwidth]{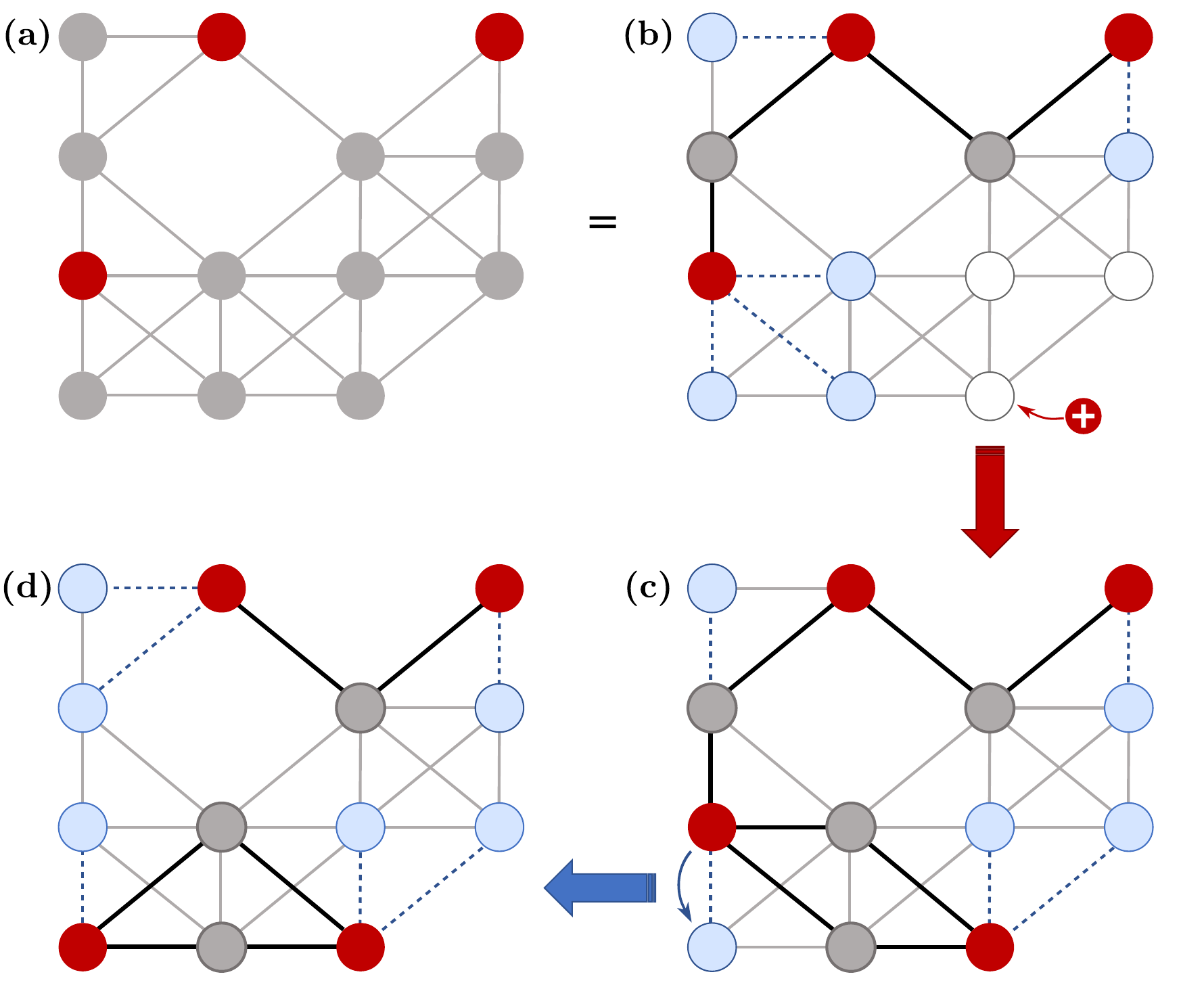}
  \caption{
    Schematic illustration of the heuristic simulated
    annealing (SA) solver. The original configuration (a) is overlaid
    with connectivity statistics in (b): nodes in the set (red),
    nodes without marked neighbors (white), and nodes with a single neighbor
    (blue).
    Potential moves are (i) removal of (red) nodes currently in the set,
    (ii) addition of currently white nodes to the set, and (iii) swapping a
    blue node with its red neighbor. 
    Grey nodes have more than one adjacent node in the set
    and no valid moves.
    From (b) to (c), one node is added to the set and the
    statistics are updated accordingly.
    From (c) to (d) one blue node is swapped with its adjacent red node.
  \label{fig:sa-pt}}
\end{figure}

Apart from the exact methods outlined above, we utilize two established
(physics-inspired) heuristic algorithms \citep{wang:15}, namely simulated annealing (SA) and parallel tempering (PT). 
For a schematic illustration see Fig.~\ref{fig:sa-pt}. In these Markov chain Monte Carlo
(MCMC) samplers, a random modification to the current solution is proposed at
each step of the algorithm and accepted depending on its effect on a specified figure
of merit. For the MIS-UD problem at hand, we use the size of the
independent set (IS) as the figure of merit to optimize. Any proposed move
(update) is then accepted with a probability governed by a temperature
parameter $T > 0$ according to the Metropolis acceptance
criterion \citep{metropolis:53}:
\begin{equation*}
  \label{metropolis}
  p_{\mathrm{accept}} =
    \begin{cases}
      1\, &\text{if } \Delta_{\mathrm{IS}} \ge 0 \\
      e^{\Delta_{\mathrm{IS}}/T}\,&\text{otherwise}.
    \end{cases}
\end{equation*}
That is, moves that \emph{increase} the size of the IS (i.e.,
$\Delta_{\mathrm{IS}}>0$) are always accepted, while those reducing its size
are suppressed -- at first only marginally at high temperatures (during initial
exploration), but then heavily at low temperatures $T$ (during final
exploitation). Our Markov dynamics consist of individual additions/removals and
swaps of neighboring sites. We ensure the independence criterion is never
violated by continuously tracking the full list of valid moves. The random
selection from this set is then biased towards adding nodes and performing
swaps in order to increase the acceptance rate (while accounting for the shift
in energy scales by adapting the cooling schedule).

\textbf{Simulated annealing} (SA) aims to find a high-quality solution by starting
from a random initial solution and initially high temperature to then gradually
lower $T\to 0$ (according to some annealing schedule) until no further
improvement is seen \citep{kirkpatrick:83}. This allows the system to first
explore the solution space while the temperature is high, but eventually drives
the state into a nearby (local, potentially global) optimum. The cooling
schedule is optimized to quickly identify this local optimum, and frequent
restarts (as specified by the parameter \codevar{num\_restarts}) from different
initial positions are used to increase the chance of finding the global
optimum.
We note that our implementation of SA differs from the one presented in Ref.~\citep{ebadi:22} in several ways: 
(i) Our implementation is not based on a (soft) penalty model as described by Eq.~(\ref{eq:hamiltonian}), but rather involves only moves compatible with the (hard) independence criterion (such that only the feasible space is searched).
(ii) Proposal probabilities are biased towards additions and exchanges to increase acceptance. 
(iii) We use a geometric cooling schedule in combination with frequent restarts, as opposed to a constant low temperature as used in Ref.~\citep{ebadi:22}. 
In particular, we note that our implementation of SA breaks detailed balance, for the sake of improved performance. 

\textbf{Parallel tempering} (PT) attempts to efficiently explore the solutions space
by simulating several Markov chains concurrently at different temperatures
\citep{swendsen:86, moreno:03, earl:05}. Exchange moves allow for swapping of
states between neighboring chains (in temperature space) such that the best
solutions are shuffled to lower temperatures for further local optimization. At
the same time, less promising candidates are moved to higher temperatures,
where large-scale restructuring is possible.

In the following we will focus on SA, since our implementation of PT did not provide any substantial performance benefits over SA for the problem instances studied here.
This is likely because of the relatively fast identification of
local minima by SA (within tens to hundreds of sweeps) compared to the mixing time
needed to exploit the benefits of PT; as such it is more efficient to restart at a random position than to invest in overcoming local energy barriers.

\section{Numerical Experiments}
\label{numerics}

\begin{figure*}
  \hspace{-4mm}
  \includegraphics[width=1.02\columnwidth]{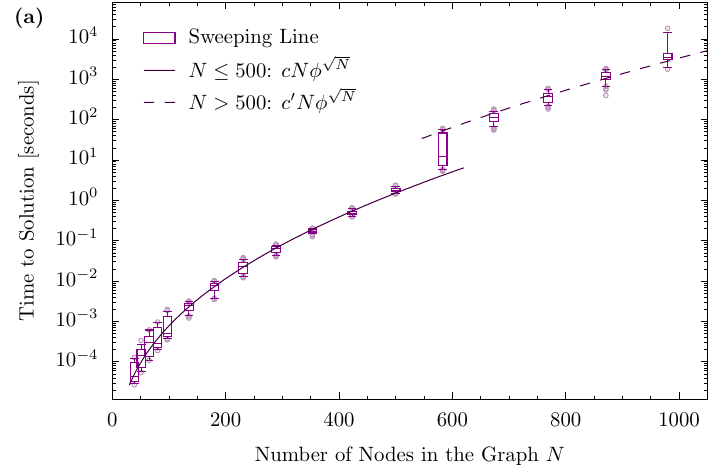}
  \hspace{4mm}
  \includegraphics[width=1.02\columnwidth]{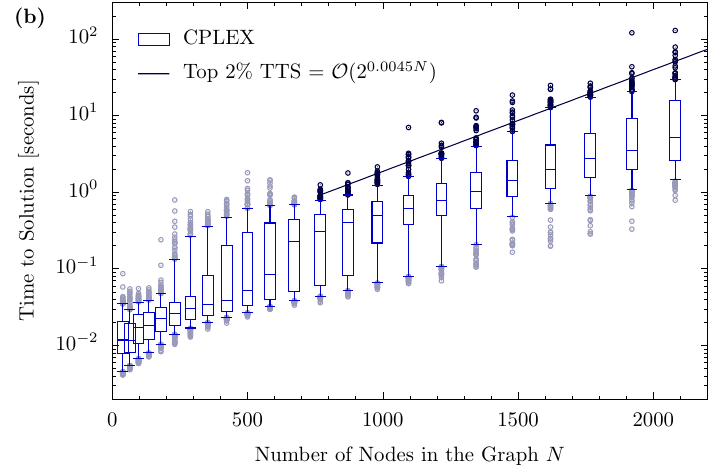}
  \caption{
    Time-to-solution (TTS) for the exact solvers.
    \textbf{(a)} TTS for the exact SLA solver as a function of system
    size $N$. For every system size $N$, 1000 random UD instances with
    $\filling=0.8$ have been considered. The data fit reasonably to
    $\mathrm{TTS}(N)\approx cN\phi^{\sqrt{N}}$, where the basis of $\phi\approx 1.62$
    is the theoretical expectation for the Fibonacci sequence.
    At larger system sizes ($N > 500$), high memory usage causes slower
    access times (cache misses), resulting in a substantially larger pre-factor $c'$.
    \textbf{(b)} TTS for the B\&B solver as a function of system
    size $N$. For every system size $N$, 1000 random UD instances have been
    considered; see Section \ref{plot-description} for box plot description.
    Problems with hundreds (thousands) of nodes can be solved to optimality in
    sub-second (minute) timescales. The solid line is the linear regression
    over instances whose TTS are in the top highest $2\%$. The linear
    regression minimizes the residual sum of squares of log(TTS).
    \label{fig:sla-bb-tts-size}
  }
\end{figure*}

We now turn to our numerical results.
We report on the TTS for all exact and heuristic
algorithms described above, as a function of system size $N$ and hardness
parameter $\HP$, with the goal to provide a comprehensive assessment of the hardness of random MIS-UD problem instances. 
For reference, we also study the MIS problem on similar yet less structured
instances (with the same number of nodes and edges), and we provide protocols to
systematically tune problem hardness (as measured by TTS) over several orders
of magnitude. The classical hardware on which our numerical experiments were
run is specified in Appendix~\ref{additional-information}. 

\subsection{Scaling with the Problem Size}

\begin{figure}
  \includegraphics[width=1.0 \columnwidth]{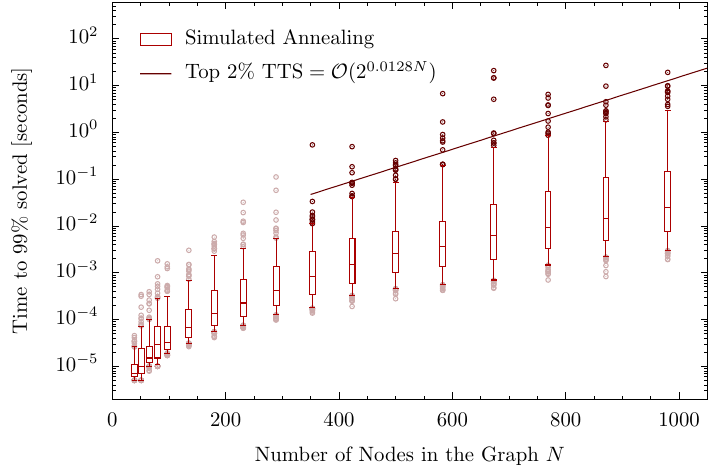}
  \caption{
    Time required to reach 99\% success probability ($\mathrm{TTS}_{99}$) for the heuristic SA
    solver as a function of system size $N$ (i.e., how long the solver
    should run for a 99\% chance of finding the optimal solution). For
    every system size $N$, 1000 random UD instances at $\filling=0.8$
    filling have been considered;
    see Section \ref{plot-description} for box plot description.
    The solid line is the linear regression
    over instances whose TTS are in the top highest $2\%$. The linear
    regression minimizes the residual sum of squares of log(TTS).
  \label{fig:pt-tts-size}}
\end{figure}

We first report on TTS for the MIS-UD
problem as a function of system size, given by the number of nodes $N \approx
\filling L^2$ at fixed density $\filling=0.8$. Note that we have run a few
additional experiments for different values of $\varrho$, with $\varrho \approx
0.8$ providing one of the hardest problems for Union-Jack connectivity (as
evidenced by the largest TTS). Thus, we primarily focus on $\filling=0.8$,
following Ref.~\citep{ebadi:22}. 
As shown in Fig.~\ref{fig:sla-bb-tts-size}, we find that the MIS-UD problem can
be solved to optimality (with both the custom SLA and the generic B\&B solvers)
for hundreds of nodes in sub-second timescales. Larger instances with up to
thousands of nodes can still be solved to optimality within minutes on
commodity hardware, without any instance-specific fine-tuning. 
For the exact SLA solver we infer a runtime scaling of
$\mathrm{TTS}_{\mathrm{SLA}}=\mathcal{O}(N\phi^{\sqrt{N}})$,
where $\phi \approx 1.62$ is the golden ratio (expected for the
scaling of the Fibonacci sequence) and the notable $\sqrt{N}$ dependence
in the exponent is attained at the expense of exponential memory requirements
(as discussed above).
For the B\&B solver we obtain
$\mathrm{TTS}_{\mathrm{B\&B}}=\mathcal{O}(2^{0.0045N})$.
Similarly, as shown in Fig.~\ref{fig:pt-tts-size}, we find that UD
instances with hundreds of nodes (i.e., $L \leq 25$) can typically be solved
efficiently in sub-second timescales with the SA heuristic. 
For the 2\% most difficult instances we observe a scaling of $\mathrm{TTS}_{99}=\mathcal{O}(2^{0.0128 N})$ for SA. However we also observe a relatively large spread spanning several orders of magnitude in $\mathrm{TTS}_{99}$ (in particular when compared to the results obtained with SLA), thus motivating a more detailed analysis of problem hardness, as discussed next.

\subsection{Scaling with the Hardness Parameter}
\label{sec:scaling-HP}

\begin{figure*}
  \hspace{-4mm}
  \includegraphics[width=1.02\columnwidth]{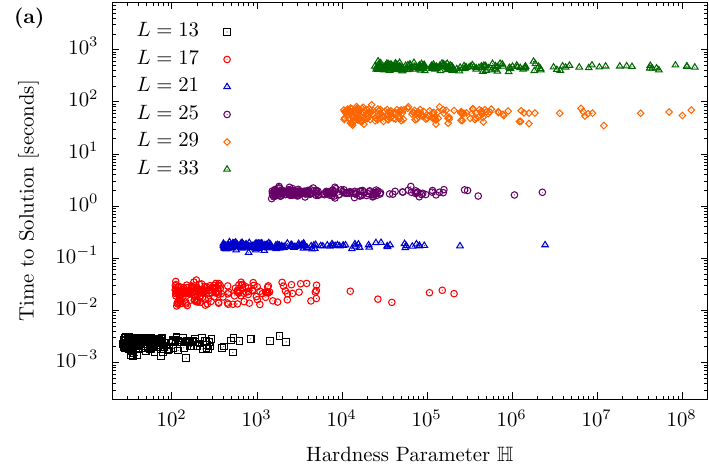}
  \hspace{4mm}
  \includegraphics[width=1.02\columnwidth]{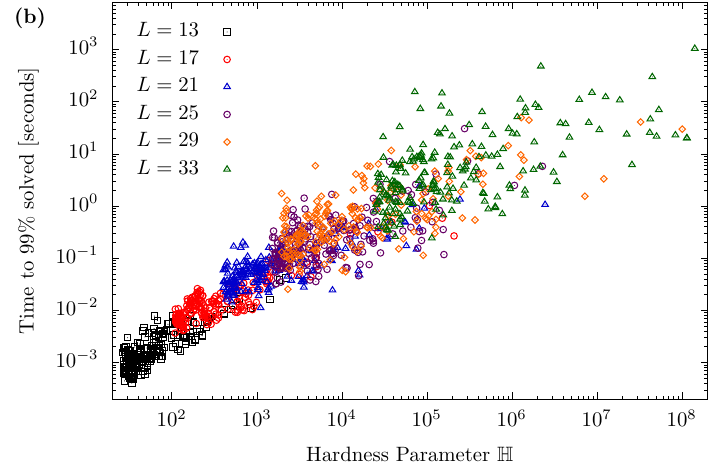}
  \caption{
    \textbf{(a)} Time-to-solution (TTS) for the exact SLA solver as a
    function of the hardness parameter $\HP$. Virtually no dependence on $\HP$
    is observed, showing that TTS is fully determined by the system size $N
    \sim L^2$.  
    \textbf{(b)} Conversely, for the Markov-chain based SA solver,
    $\mathrm{TTS}_{99}$
    shows a strong correlation with the hardness parameter $\HP$, as expected. 
  \label{fig:sla-tts-hp}}
\end{figure*}

Some of the results presented
in the previous section display significant instance-to-instance variations in $\mathrm{TTS}_{99}$, potentially
spanning several orders of magnitude, even for fixed system size $N$. 
As argued in Ref.~\citep{ebadi:22}, these variations may largely be due to
(large) differences in the total number of MIS solutions, given by the ground
state degeneracy $D_{\mathrm{MIS}}$, a quantity that can be calculated either
with tensor-network methods \citep{liu:22} or within the exact SLA method
outlined above (at least for small to intermediate system sizes up to $N \sim
1000$). 
Intuitively, the less degenerate the ground state, the harder it is to hit
the global optimum.
In particular, based on experiments with up to $N=80$ qubits in
Ref.~\citep{ebadi:22}, a quantum speedup over classical SA has been reported in
the dependence of the success probability $P_{\mathrm{MIS}}$ on the hardness
parameter $\HP$ that accounts for both the degeneracy of the ground as well as
first excited states (denoted by $D_{\mathrm{MIS}}$ and $D_{\mathrm{MIS}-1}$,
respectively), as given in Eq.~(\ref{eq:hardness}). 

We now follow Ref.~\citep{ebadi:22} and consider algorithmic performance in
terms of this hardness parameter $\HP$ for values as high as $\HP \sim 10^8$,
complementing existing results based on classical SA \citep{ebadi:22} with
results for the SLA, and B\&B solvers. Note that hyperparameter
optimization has been performed for our heuristic solvers, although without any
instance-to-instance fine-tuning.
For direct comparison, the results for the exact SLA as well as the heuristic
SA solvers are displayed in Fig.~\ref{fig:sla-tts-hp} showing a
remarkably different behavior.    
Qualitatively, we find that $\mathrm{TTS}_{99}$ for the SA solver displays a strong dependence
on the hardness parameter $\HP$, in line with results reported in
Ref.~\citep{ebadi:22}. Conversely, virtually no dependence between TTS and $\HP$ is observed for
the exact SLA solver, as expected, thereby demonstrating that the conductance-like
hardness parameter $\HP$ successfully captures hardness for algorithms
undergoing Markov-chain dynamics. 
Alternative algorithmic paradigms such
as sweeping line or branch and bound, however, may require a different notion of
hardness.
Similarly, for the B\&B solvers we find that TTS is (weakly) correlated with
the hardness parameter $\HP$, although mostly because of their common
correlation with the system size $N$. Specifically, using linear regression, we
have found that the partial correlation of $\log_{10}(\HP)$ and $\log_{10}(\mathrm{TTS})$
(controlling for system size) is smaller than 0.05; see Appendix \ref{correlation-TTS} for further details. This weak correlation suggests that (similarly to our SLA results) the hardness parameter $\HP$ does not appear
to be a reliable measure of hardness for B\&B-type solvers.

\begin{figure}
  \hspace{-4mm}\includegraphics[width=1.02\columnwidth]{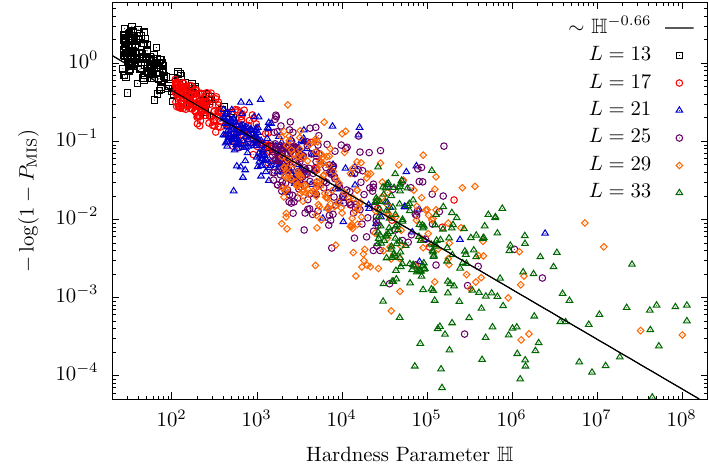}
  \caption{
    Estimated success probability $P_{\mathrm{MIS}}$ for the heuristic SA
    solver as a function of the hardness parameter $\HP$. Here we plot
    $-\log(1-P_{\mathrm{MIS}})$ for SA with a fixed depth of 32, for UD graphs
    selected from the top two percentile of hardness parameter $\HP$ for each
    system size $L=13, \dots, 33$. Hollow points represent our SA implementation
    with bias moves in valid configuration space. 
    Power-law fits to the form $\sim
    \HP^{-\alpha}$ are used to extract scaling performance with graph hardness
    $\HP$.
    \label{fig:sa-pmis-hp-fits}
  }
\end{figure}

Finally, we complement the TTS results above with results
for the success probability $P_{\mathrm{MIS}}$ as a function of the
hardness parameter $\HP$ for the SA solver, as done in Ref.~\citep{ebadi:22}
for both SA and quantum algorithms for instances with hardness of up to $\HP
\sim 10^3$. 
Our results with hardness of up to $\HP \sim 10^8$ are shown in
Fig.~\ref{fig:sa-pmis-hp-fits}. 
Following Ref.~\citep{ebadi:22}, for fixed depth (i.e., number of SA sweeps),
fits are provided assuming the functional form $P_{\mathrm{MIS}} = 1 - \exp(-C
\HP^{-\alpha})$, where $C$ refers to a positive fitted constant that (in
general) could have polynomial dependence on the system size $N$, and smaller
values of $\alpha$ yield larger success rate $P_{\mathrm{MIS}}$. 
While there are significant variations in the data, on average we observe a scaling
$P_{\mathrm{MIS}} \approx 1 - \exp(-C \HP^{-0.66})$ (solid black line), i.e.,
$\alpha=0.66$; additional results with size-dependent depth (not shown) suggest
an even higher success probability $P_{\mathrm{MIS}}$ but preclude simple fits
because of additional size-dependent effects in the data. 
In particular, a fixed depth of 32 is arguably too small for the largest
instances considered here, but was chosen nevertheless for better comparison
with results reported in Ref.~\citep{ebadi:22}. 
For comparison, a fit with the exponent $\alpha=1.03$ was reported for SA in Ref.~\citep{ebadi:22}.
If one restricts the analysis to graphs with minimum energy gaps sufficiently
large to be resolved in the duration of the (noisy) quantum evolution, the
optimized quantum algorithm demonstrated in Ref.~\citep{ebadi:22} was shown to
fit best to $\alpha=0.63$, i.e., comparable to $\alpha=0.66$ as 
found with our implementation of classical SA.

\subsection{Beyond Union-Jack Connectivity}

\begin{figure}
  \hspace{-4mm}\includegraphics[width=1.02\columnwidth]{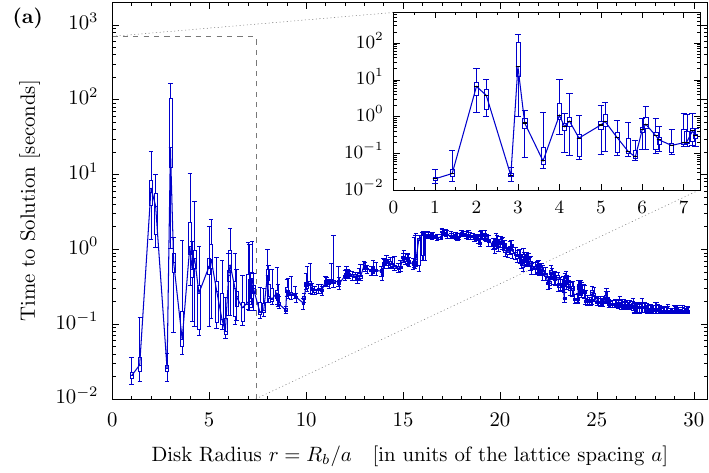}\\
  \vspace{3mm}
  \hspace{-4mm}\includegraphics[width=1.02\columnwidth]{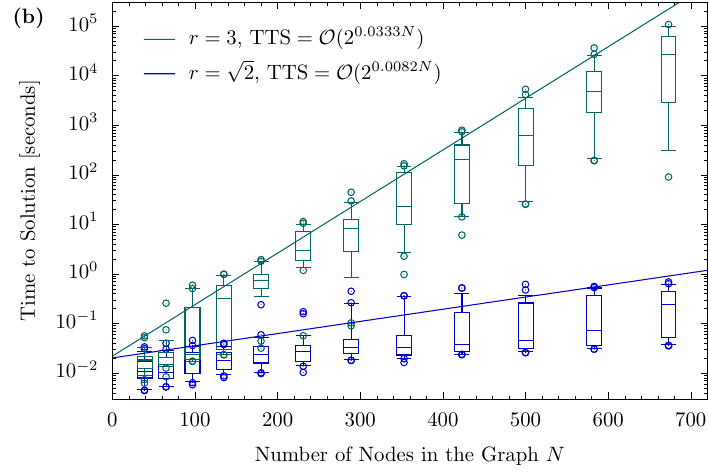}\\
  \vspace{-1mm}
  \caption{ 
    \textbf{(a)} Hardness transition as a function of the disk radius (in units of the lattice spacing) $r=R_{b}/a$, as given by the time-to-solution (TTS) for the B\&B solver, shown here for system size $L=21$ and density $\filling=0.8$ (i.e., $N\approx 350$), with 100 random seeds per radius.
    \textbf{(b)}: TTS as a function of system size $N=\filling L^2$ for $r=\sqrt{2}$ (blue) and $r=3$ (green), the latter referring to the pronounced peak observed in (a).  
    The solid lines are a linear regression fit over 100 instances with TTS in the highest $10\%$, with corresponding $R^2$ values of 0.87 and 0.98 for $r=\sqrt{2}$ and $r=3$, respectively. Instances with $r=3$ appear to be much harder than those with $r=\sqrt{2}$. See Section \ref{plot-description} for box plot description.
  \label{fig:bb-tts-radius}}
\end{figure}

\begin{figure}
  \hspace{-4mm}\includegraphics[width=1.02\columnwidth]{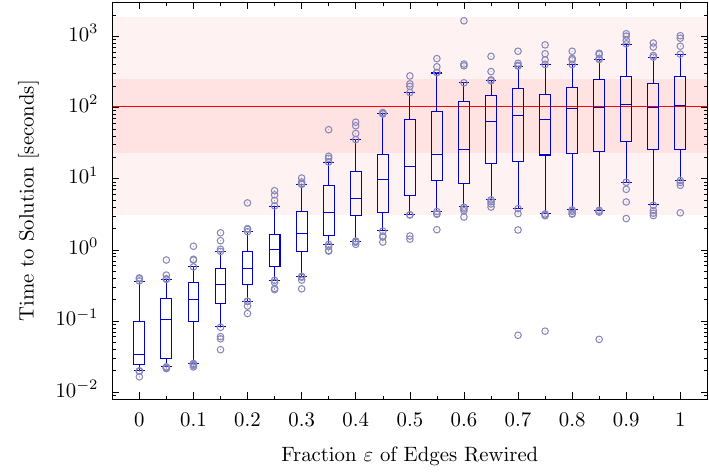}
  \vspace{-1mm}
  \caption{
    Hardness transition from unit-disk (UD) to random Erd\H{o}s-R\'enyi (ER) graphs. 
    Time-to-solution (TTS) for MIS as a function of the fraction $\varepsilon$ of edges rewired, with 150 random seeds. Starting from Union-Jack-type UD graphs (left), edges are randomly selected and rewired, thereby gradually breaking the UD connectivity, and ultimately generating random ER graphs (right).
    While the original UD graphs can be solved to optimality in $\sim 10^{-2}\mathrm{s}$, comparable ER graphs (with the same number of nodes and edges) display a TTS orders of magnitudes larger. The red line and the two shaded areas refer to the median TTS over 500 instances for MIS on random ER graphs, the TTS among the $25\%$ and $75\%$ and the minimum and maximum whiskers, respectively. Numerical parameters: $L=21$, and $\filling=80\%$.  
  \label{fig:breaking_udg}}
\end{figure}

To provide further context for the results reported above, we now study hardness as we gradually change the topology of the problem instances. Specifically, going beyond Union-Jack-type instances (with $\sqrt{2}\leq r < 2$ fixed) as studied so far, we analyze TTS following two protocols by either (i) systematically tuning the blockade radius $r=R_{b}/a$ or (ii) randomly rewiring edges of the graph. While protocol (i) prepares UD graphs (with varying connectivity), protocol (ii) explicitly breaks the UD structure via random (potentially long-range) interactions, ultimately preparing random (structure-less) ER graphs. The results of this analysis are shown in Figs.~\ref{fig:bb-tts-radius} and \ref{fig:breaking_udg}, respectively.  
We find that problem hardness (as measured here by TTS for the established B\&B solver) can be tuned systematically over several orders of magnitude. 

As shown in Fig.~\ref{fig:bb-tts-radius}, we find that the MIS-UD problem is relatively easy for both small ($r<2$) and large radii (as expected because the MIS problem is trivial for both edge-less and complete graphs), but significantly harder in between, with a pronounced peak at $r=3$. For example, for $L=21$ and $\filling=0.8$ (giving $N\approx 350)$ we observe an increase in TTS over $2-3$ orders of magnitude for $r=3$ instances compared to instances with small or large radii. 
This behavior appears to be generic, as we have observed similar behavior with the SLA solver (cf. Fig.~\ref{fig:sla-tts-radius}), again showing pronounced peaks in TTS for $r=2, 3, 4$; see Appendix \ref{additional-numerics} for more details. 
This observation may be attributed to a density-of-states-like effect: While the boundary size grows monotonically with $r$, we find that the number of boundary variants peak at $r=2, 3, 4$ before decreasing with $r$ as the (feasible) state space becomes smaller, thus correlating the observed TTS behavior with the number of boundary variants. 
This spread in TTS is found to increase with system size, as shown in Fig.~\ref{fig:bb-tts-radius} (b) for instances with Union-Jack topology (with $r=\sqrt{2}$) as well as instances with $r=3$.  

Alternatively, following protocol (ii) the MIS problem appears to become orders of magnitude harder when randomly rewiring edges (thereby breaking the UD structure), as shown in Fig.~\ref{fig:breaking_udg}. 
In particular, we find that structure-less ER graphs can yield a TTS orders of magnitudes larger than more structured UD graphs (with the same average number of nodes and edges), in agreement with similar results for random UD graphs where vertices are placed randomly (and not on a square lattice) in a two-dimensional box with some fixed density \citep{pichler:18}.

\subsection{Implementation with Rydberg Arrays} 

The two protocols outlined above may be implemented in future experiments with Rydberg atom arrays, either by (i) tuning the Rydberg blockade radius $R_{b}$ and/or the lattice spacing $a$, or by (ii) implementing embedding schemes with ancilla Rydberg chains, as proposed in Refs.~\citep{nguyen:23, kim:22}, thus suggesting a potential recipe to benchmark quantum algorithms on instances orders of magnitude harder (for established classical solvers) than previously studied. 
For example, for today's Rydberg atom arrays (such as the QuEra Aquila device available through Amazon Braket), we estimate that $R_{b}/a \sim 3$ (amounting to a maximum degree of $d_{\mathrm{max}} \sim 28$) should be achievable already today, with potentially even larger values enabled by future hardware improvements. 
Experiments like these may also provide new insights into effects stemming from the long-range interaction tails associated with the Rydberg interactions. 

\subsection{Prospects for Quantum Speedups}

With the goal to help identify regimes and system sizes where quantum algorithms could be useful, we now briefly revisit our results in the light of on-going efforts towards quantum advantage. 
Adopting the taxonomy put forward in Ref.~\citep{mandra:16}, within a larger hierarchy of potential quantum speedups, the quantum speedup demonstrated in Ref.~\citep{ebadi:22} could be classified as \textit{limited sequential quantum speedup}, as it was obtained by comparing a quantum annealing type algorithm over a particular implementation of the classical (sequential) simulated annealing algorithm (that was designed to fulfill detailed balance).
Here, we have tried to extend the classical SA benchmarking results (by pushing the hardness parameter up to $\HP \sim 10^8$), with a different implementation of SA that breaks detailed balance but shows better performance. 
While our SA-based scaling results hint at performance similar to the quantum algorithm's performance in Ref.~\citep{ebadi:22}, we note that the corresponding exponent shows a dependence on the somewhat arbitrary cut-off in hardness and additional dependence on system size $N$ if the depth is not fixed; the details thereof will be analyzed in future research.
Still, within the aforementioned hierarchy of quantum speedups, our analysis points at a potential next milestone, in the form of the experimental demonstration of a (more general) \textit{limited non-tailored quantum speedup}, by (for example) comparing the performance of the quantum algorithm to the best-known generic classical optimization algorithm. 
In particular, B\&B solvers (as studied here) could be good candidates to fill the role for the latter, and the protocols outlined above motivate potential quantum experiments for such studies. 
Again, for reference for the MIS-UD problem on Union-Jack instances, empirically for most instances (taken as 98\%-th percentile) we can upper-bound the TTS needed by classical B\&B solvers through a runtime scaling of $\mathrm{TTS} = \OO(1.0031^N)$ (c.f. Fig.~\ref{fig:sla-bb-tts-size}), setting an interesting, putative bar for quantum algorithms to beat. 

\section{Conclusion and Outlook}
\label{conclusion}

In summary, we have studied the maximum independent set problem on unit-disk graphs (as it can be encoded efficiently with Rydberg atom arrays \citep{ebadi:22, pichler:18}), using a plethora of exact and heuristic classical algorithms. 
We have assessed problem hardness, showing that instances with thousands of nodes can be solved to optimality within minutes using both custom and generic commercial solvers on commodity hardware, without any instance-specific fine-tuning. 
We have also performed a detailed scaling analysis, showing that our implementation of classical simulated annealing is competitive with the quantum algorithm's
performance in Ref.~\citep{ebadi:22}.
Finally, we have devised protocols to systematically tune problem hardness over several orders of magnitude. 
In future studies, the problem hardness may be tuned even further by generalizing our work from unweighted to weighted graphs, thereby potentially lifting the ground state degeneracy, as could be studied with local detuning control in Rydberg atom arrays. 
We hope that these protocols may trigger interesting future experiments further exploring the hardness of the MIS problem with Rydberg atom arrays as pioneered in Ref.~\citep{ebadi:22}.  

{\bf Code availability statement}: An open source demo version of our code is publicly available at \texttt{github.com/jpmorganchase/hardness-of-mis-on-udg}.

\begin{acknowledgments}
We thank Maddie Cain, Alex Keesling, Eric Kessler, Harry Levine, Mikhail Lukin, Hannes Pichler, and Shengtao Wang for fruitful discussions. 
We thank Peter Komar and Gili Rosenberg for detailed reviews of the manuscript,
and Marjan Bagheri, Victor Bocking, Alexander Buts, Lou Romano, Peter Sceusa, and Tyler Takeshita for their support.
\end{acknowledgments}

\section*{Disclaimer}
This paper was prepared for informational purposes with contributions from the Global Technology Applied Research center of JPMorgan Chase \& Co. This paper is not a product of the Research Department of JPMorgan Chase \& Co. or its affiliates. Neither JPMorgan Chase \& Co. nor any of its affiliates makes any explicit or implied representation or warranty and none of them accept any liability in connection with this paper, including, without limitation, with respect to the completeness, accuracy, or reliability of the information contained herein and the potential legal, compliance, tax, or accounting effects thereof. This document is not intended as investment research or investment advice, or as a recommendation, offer, or solicitation for the purchase or sale of any security, financial instrument, financial product or service, or to be used in any way for evaluating the merits of participating in any transaction.

\appendix

\section{Additional Information}
\label{additional-information}

In this Appendix we provide further details for selected aspects discussed in the main text. 
In section \ref{number-edges} we derive the maximum and expected number of edges for Union-Jack-like graphs with density $\filling$. 
In section \ref{classical-hardware} we detail the hardware on which our numerical experiments were run. 
In section \ref{plot-description} we provide additional information on our plots. 
In section \ref{tts-hardness-scaling} we derive the scaling of TTS with the hardness parameter $\HP$ for SA. 

\subsection{Expected Number of Edges in UD Graphs}
\label{number-edges}

In this section we show that for quasi-planar UD graphs with Union-Jack-type connectivity and filling fraction $\filling$ (as considered here) the expected (or average) number of edges amounts to $\overline{|\mathcal{E}|} = \filling^2 (4 L^2 - 6 L + 2)$; for full filling of the square lattice at $\filling=100\%$ we recover $\overline{|\mathcal{E}|} \rightarrow |\mathcal{E}|_{\mathrm{max}}=4 L^2 - 6 L + 2$. 
To this end we represent nodes as independent random variables $X_k$, following a Bernoulli distribution, with value $X_k=1$ if the node is present with probability $\filling$, and $X_k=0$ otherwise (if the node is absent). 
The probability that an edge $(i,j)$ from $\mathcal{E}_{\mathrm{max}}$ exists in $\mathcal{E}$ is then given by $E[X_iX_j]=\filling^2$. 
As such, the expected number of edges in $\mathcal{E}$ follows as  $E[\sum_{(i,j)\in\mathcal{E}_{\mathrm{max}}} X_iX_j]=\filling^2 (4 L^2 - 6 L + 2)$.
We have used this relation to generate random ER graphs $\mathcal{G}(n,p)$ with desired (average) edge probability $p=\overline{|\mathcal{E}|} / (n(n-1)/2)$; we note that the variance of $|\mathcal{E}|$ differs between the ER graphs and the UD graphs.

\subsection{Classical Hardware}
\label{classical-hardware}

In this section we specify the classical hardware on which our numerical experiments were run. All B\&B-based results obtained with optimization by CPLEX 20.1 were collected from executions of this software using the Python package docplex with default parameters. 
The hardware employed is an Intel(R) Xeon(R) Platinum 8259CL CPU @ 2.50GHz with 8 cores.
For numerical experiments shown in Fig.~\ref{fig:sla-bb-tts-size}(b), the median overhead due to multi-threading was found to be nearly constant over the entire range of problem sizes. As such, the $\mathcal{O}$ complexity is insensitive to the maximum number of threads used by CPLEX. 
The results for the sweeping line algorithm and simulated annealing, where time-to-solution was considered, were performed on an AMD(R) Ryzen(R) 9 5950X @ 4.90Ghz. Various hardware was employed in sampling instances for hardness and $P_{\mathrm{MIS}}$ (where the runtime is irrelevant).

\subsection{Plots Description}
\label{plot-description}

If not stated otherwise, boxes in box plots correspond to the 16\% and 84\% percentiles
of the $\log_{10}$ variable plotted, and the whiskers refer to the the 2\% and 98\% percentiles. The horizontal line within the box denotes the median value.
Points outside the whiskers are plotted individually. This was done to highlight the
top 2\% which are used when analyzing the scaling behavior for hard instances.

\subsection{Scaling of TTS with Hardness}
\label{tts-hardness-scaling}

In this section we derive the expected scaling of $\mathrm{TTS}_{99}$ with the hardness parameter $\HP$ for SA. 
In the main text, for fixed depth we have provided fits assuming the functional form $P_{\mathrm{MIS}} = 1 - \exp(-C \HP^{-\alpha})$; see Fig.~\ref{fig:sa-pmis-hp-fits}. 
Here, we now focus on hard instances where $\HP$ is large and $P_{\mathrm{MIS}}$ is small. 
For sufficiently large values of $\HP$, we can approximate $P_{\mathrm{MIS}} \approx C \HP^{-\alpha}$, giving $P_{\mathrm{MIS}} \sim \HP^{-\alpha}$. 
For small values of $P_{\mathrm{MIS}}$ we have $\mathrm{TTS}_{99} \sim 1/P_{\mathrm{MIS}}$, yielding the scaling 
\begin{equation}
\mathrm{TTS}_{99} \sim \HP^{\alpha}.
\end{equation}
Thus, for large hardness we expect TTS to scale with the same exponent $\alpha$ as found from the fit shown in Fig.~\ref{fig:sa-pmis-hp-fits}.
We have further verified this result numerically by excluding small systems with small hardness. 

\section{Additional Numerical Results}
\label{additional-numerics}

In this Appendix we present additional numerical results, complementing the results shown in the main text. 
In section \ref{filling} we provide additional results for problem hardness as a function of the filling fraction $\filling$. 
In section \ref{correlation-TTS} we analyze a potential correlation between the hardness parameter $\HP$ and TTS for our B\&B solvers. 
In section \ref{large-radius} we provide additional results for problem hardness as a function of the unit-disk radius $r$. 

\subsection{Filling Fraction}
\label{filling}

\begin{figure}
  \hspace{-4mm}\includegraphics[width=1.02\columnwidth]{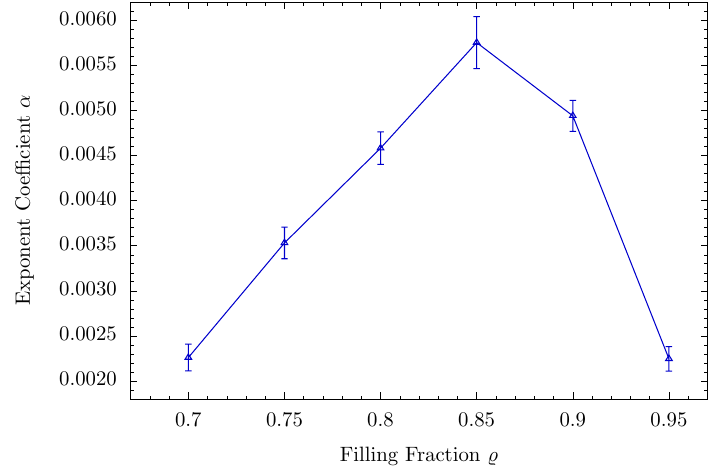}
  \caption{Coefficient $\alpha$ extracted from time-to-solution (TTS) scaling with system size as $\mathrm{TTS}=\mathcal{O}(2^{\alpha N})$ for the B\&B solver as a function of the lattice filling $\filling$. In the main text we have focused on $\filling=0.8$. All results have been taken from the top 2\% TTS instances over 1000 random instances. Error bars correspond to the $95\%$ confidence interval.
  \label{fig:bb-tts-density}}
\end{figure}
 
In the main text we have focused on instances with Union-Jack-like connectivity and filling fraction $\filling=80\%$, following Ref.~\citep{ebadi:22}. 
In Fig.~\ref{fig:bb-tts-density} we provide additional results for problem hardness as a function of the density $\filling \in [0.7, 0.9]$, given in terms of the coefficient $\alpha$ extracted from TTS scaling with system size as $\mathrm{TTS}=\mathcal{O}(2^{\alpha N})$ for the B\&B solver.
We find that instances with $\filling \approx 80\%$ are among the hardest instances, with the hardness maximum around $\filling \approx 85\%$.

\subsection{Correlation of Hardness with TTS for B\&B Solvers}
\label{correlation-TTS}

In this section we analyze a potential correlation between the hardness parameter $\HP$ and TTS for our B\&B solvers. 

\begin{figure}
  \hspace{-1mm}\includegraphics[width=1.01 \columnwidth]{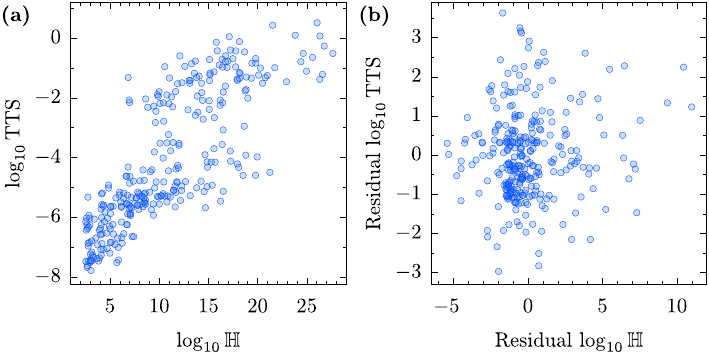}
  \caption{Correlation analysis for B\&B solver. Results are shown for the hardest $2\%$ according to the hardness parameter $\HP$ on 1,000 instances for each (odd) problem size from $L=7$ to $35$. \textbf{(a)} Scatter plot of hardness $\HP$ and TTS for B\&B solver. The corresponding Pearson correlation amounts to 0.81. \textbf{(b)} Scatter plot of the residuals from the linear regression of $\HP$ and TTS with system size $N$. Here the Pearson correlation is found to be 0.01.} 
  \label{fig:HP-TTS-correlation}
\end{figure}

At first, we observe a correlation between the logarithms of $\HP$ and TTS with a Pearson correlation reaching 0.48. 
However, when removing a linear dependence with the system size $N$ to both of these quantities, the correlation is found to drop to 0.044. Selecting only the hardest instances (as given by the top $2\%$ of hardness $\HP$) exacerbates this phenomenon as shown in Fig. \ref{fig:HP-TTS-correlation}.
Note that the linear regression is done without a minimum size threshold 
because the hardness parameter $\HP$ becomes rapidly prohibitively long to evaluate.

\subsection{Additional Results for Large-Radius Instances}
\label{large-radius}

In this section we provide additional results for problem hardness as a function of unit-disk radius $r$ for random UD instances. 

\begin{figure}
  \vspace{1mm} 
  \hspace{-2mm}\includegraphics[width=1.02\columnwidth]{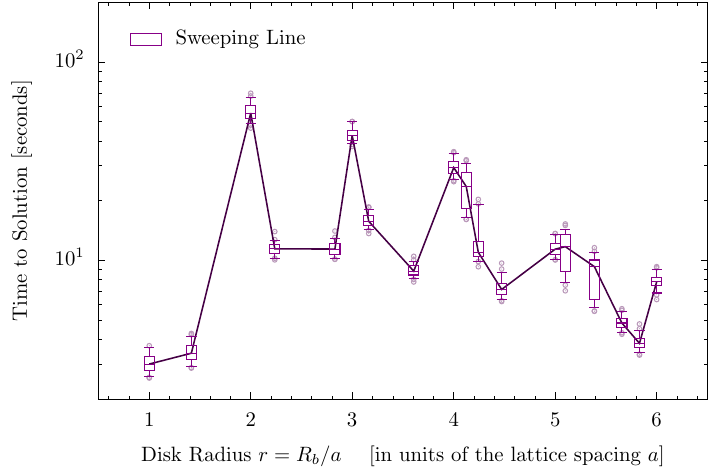}
  \caption{Time-to-solution (TTS) for the SLA solver as a function of the disk radius $r$ for random UD instances, for system size $L=21$ and density $\filling=0.8$ (i.e., $N \approx 350$), with 100 random seeds per radius. Similar to our results for our B\&B solver, we observe distinct peaks at $r=2,3,4$. 
  \label{fig:sla-tts-radius}}
\end{figure}

First we provide results for the exact SLA solver. 
Our results for TTS as a function of the disk radius $r$ for random UD instances are shown in Fig.~\ref{fig:sla-tts-radius}. 
Similar to our results for our B\&B solver, we observe distinct peaks at $r=2,3,4$, thus further motivating future experiments with Rydberg atom arrays on these instances. 

\begin{figure}
  \hspace{-2mm}\includegraphics[width=1.02 \columnwidth]{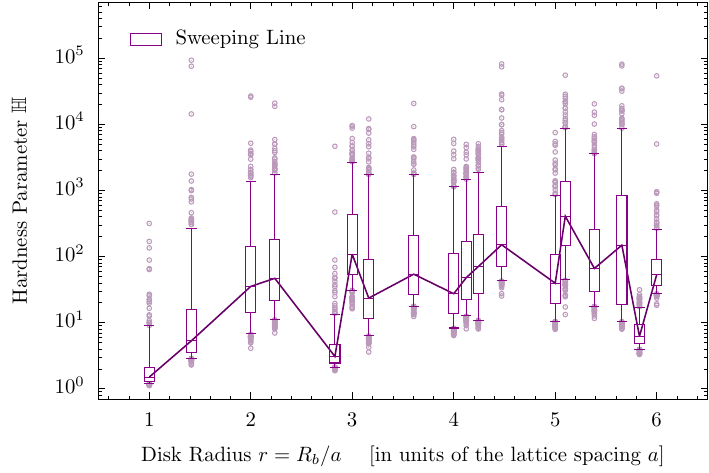}
  \caption{Hardness parameter $\HP$ as a function of the radius of the unit-disk graph (UDG), for system size $L=21$ and density $\filling=0.8$ (i.e., $N\approx 350$), with 1000 random seeds per radius.  
  \label{fig:hp-radius}}
  \vspace{-1mm}
\end{figure}

We also analyze the dependence of the hardness parameter $\HP$ [as defined in Eq.~(\ref{eq:hardness})] on the unit-disk radius $r$, knowing that $\HP$ largely determines problem hardness and thus algorithmic performance for Markov-chain based algorithms such as SA as well as the hybrid quantum algorithm in Ref.~\citep{ebadi:22}. 
The results of this analysis are displayed in Fig.~\ref{fig:hp-radius}. 
While we do observe some dependence, we do not observe very pronounced peaks as seen for the B\&B and SLA solvers in Figs.~\ref{fig:bb-tts-radius} and \ref{fig:sla-tts-radius}, respectively. 
Again, this observation motivates future experiments with Rydberg atom arrays,  
as the likelihood for a potential quantum speedup may be larger on these instances, 
thus potentially helping to identify new regimes where quantum algorithms can be useful.

\bibliography{bibliography}

\end{document}